\documentclass[pre,showpacs,notitlepage,twocolumn]{revtex4-1}
\pdfoutput=1

\usepackage[colorlinks,linkcolor={blue},citecolor={blue},urlcolor={blue}]{hyperref}

\usepackage{graphicx}

\usepackage{bm,amsmath}

\usepackage{latexsym}

\usepackage{verbatim}

\usepackage{color}

\usepackage{subfigure}

\usepackage{gensymb}

\usepackage{soul}

\usepackage[dvipsnames]{xcolor}

\newcommand{\nhat}{\hat{\bf n}}

\newcommand{\xhat}{\hat{\bf x}}

\newcommand{\nb}{\mathbf{n}}

\newcommand{\rhat}{\hat{\bf r}}

\newcommand{\rb}{{\bf r}}
\newcommand{\ub}{{\bf u}}

\newcommand{\Fb}{{\bf F}}
\newcommand{\fb}{{\bf f}}
\newcommand{\Hb}{{\bf H}}
\newcommand{\Ib}{{\bf I}}

\newcommand{\Tb}{{\bf T}}

\newcommand{\Mb}{{\bf M}}
\newcommand{\Lb}{{\bf L}}

\newcommand{\Ob}{{\bf O}}

\newcommand{\vb}{{\bf v}}

\newcommand{\xb}{{\bf x}}

\newcommand{\omegab}{\mbox{\boldmath $\omega$\unboldmath}}

\newcommand{\mutens}{\mbox{\boldmath $\mu$\unboldmath}}

\newcommand{\nablab}{\mbox{\boldmath $\nabla$\unboldmath}}

\newcommand{\beq}{\begin{equation}}

\newcommand{\eeq}{\end{equation}}

\newcommand{\bea}{\begin{eqnarray}}

\newcommand{\eea}{\end{eqnarray}}

\newcommand{\tca}[1]{\textcolor{black}{#1}}

\begin{document}

\title{Flow properties  and hydrodynamic interactions of rigid spherical microswimmers }

\author{Tapan Chandra Adhyapak}
\email{tadhyapa@uni-mainz.de}
\affiliation{Institut f\"{u}r Physik, Johannes Gutenberg-Universit\"{a}t
Mainz, Staudingerweg 7-9, 55128 Mainz, Germany}

\author{Sara Jabbari-Farouji}
\email{sjabbari@uni-mainz.de}
\affiliation{Institut f\"{u}r Physik, Johannes Gutenberg-Universit\"{a}t
Mainz, Staudingerweg 7-9, 55128 Mainz, Germany}

\date{\today}

\begin{abstract}

We analyze a minimal model for a rigid spherical microswimmer and explore the
consequences of its extended surface on the interplay between its
self-propulsion and flow properties. The model is the first order
representation of microswimmers, such as  bacteria and algae, with rigid bodies
and flexible propelling appendages. The flow field of such a microswimmer 
\tca{at finite distances}
significantly differs from that of a point-force (Stokeslet) dipole. 
For a suspension of microswimmers, we  derive  the grand mobility
matrix that connects the motion of an individual swimmer to the active and
passive forces and torques acting on all the swimmers.  Our investigation of
the mobility tensors reveals that 
hydrodynamic interactions among rigid-bodied microswimmers 
\tca{differ considerably from}
those among  the corresponding point-force dipoles.  Our results are relevant
for the study of collective behavior of hydrodynamically interacting
microswimmers by means of Stokesian dynamics simulations at moderate
concentrations.

\end{abstract}


\maketitle



\section{Introduction}

Microswimmers have attracted a lot of attention from soft matter physicists
over the last years \cite{lauga2009, bacterial_turb2, aranson2011,
active_superfluid, bacteria_LC1, bacteria_LC2}.  Intensive research has been
directed towards living microswimmers such as bacteria \cite{tapan_PRE2015,
reinhardPRL2013, grahamPRE2011, gomperSM2012, saragosti2011, strong,
turner2007, wada2008, pohl_2017, larson2015}, algae
\cite{rafai_viscosity_PRL2010, friedrichPRL2012} and other microorganisms
\cite{sperm_science, sperm_scatter, gompper_cilia, davod2015}. Moreover,
physicists'contributions have reached beyond just the living systems and
studies on artificial microswimmers have also been on the rise
\cite{artificial1, artificial2, artificial3}.  Microswimmers are the prime
examples of active particles \cite{SR_RevMod2013}.  Their individual dynamics
involves challenging physics \cite{squirmer_lighthill1952, powersPRL2005,
lauga_PRL2011, lauga_AnnRev2016, tapan_Softmatter2016}, and their collective
behaviors often exhibit surprises that have triggered a number of new
directions in physics per se \cite{active_superfluid, bacterial_turb1,
bacterial_turb2, aranson2011}.

Microswimmers immersed in a viscous fluid generate long-range flows as they
move.  They also experience hydrodynamic interactions as they react to the
local flow generated by the others.  Many distinctive dynamical features of
microswimmers  result from the interplay between their self-propulsion and
mutual hydrodynamic interactions \cite{rafai_viscosity_PRL2010, stocker2012,
bacterial_turb2, andreas_PRL2014}.  However, the consequences of hydrodynamic
interactions on their dynamics are not fully understood.  Although the
importance of  the particle size and shape on the hydrodynamics  of  passive
objects is well-known \cite{dhont, kim_karilla}, the influence of the finite
body-size on the swimming behavior of active particles has received little
attention \cite{graham_dumbbell_PRL2005, active_rotor_2012, menzel2016}.

Most  studies of collective behavior treat microswimmers as hydrodynamic
point-force (Stokeslet) dipoles   \cite{lauga2009, runtumble_hydro,
graham_dumbbell_PRL2005}. This approximation accounts for the hydrodynamic
interactions correctly at very dilute concentrations where the body-size of the
individual swimmers is negligible with respect to the  interparticle
separations. Here, we analyze a minimal model for a rigid spherical
microswimmer by incorporating the influence of its extended surface on the flow
created during the self-propulsion. We thereby investigate the role of the
finite-body size on its locomotion, flow properties and on the hydrodynamic
interactions with other swimmers.

\begin{figure}
\begin{center}
\includegraphics[width=.8\columnwidth]{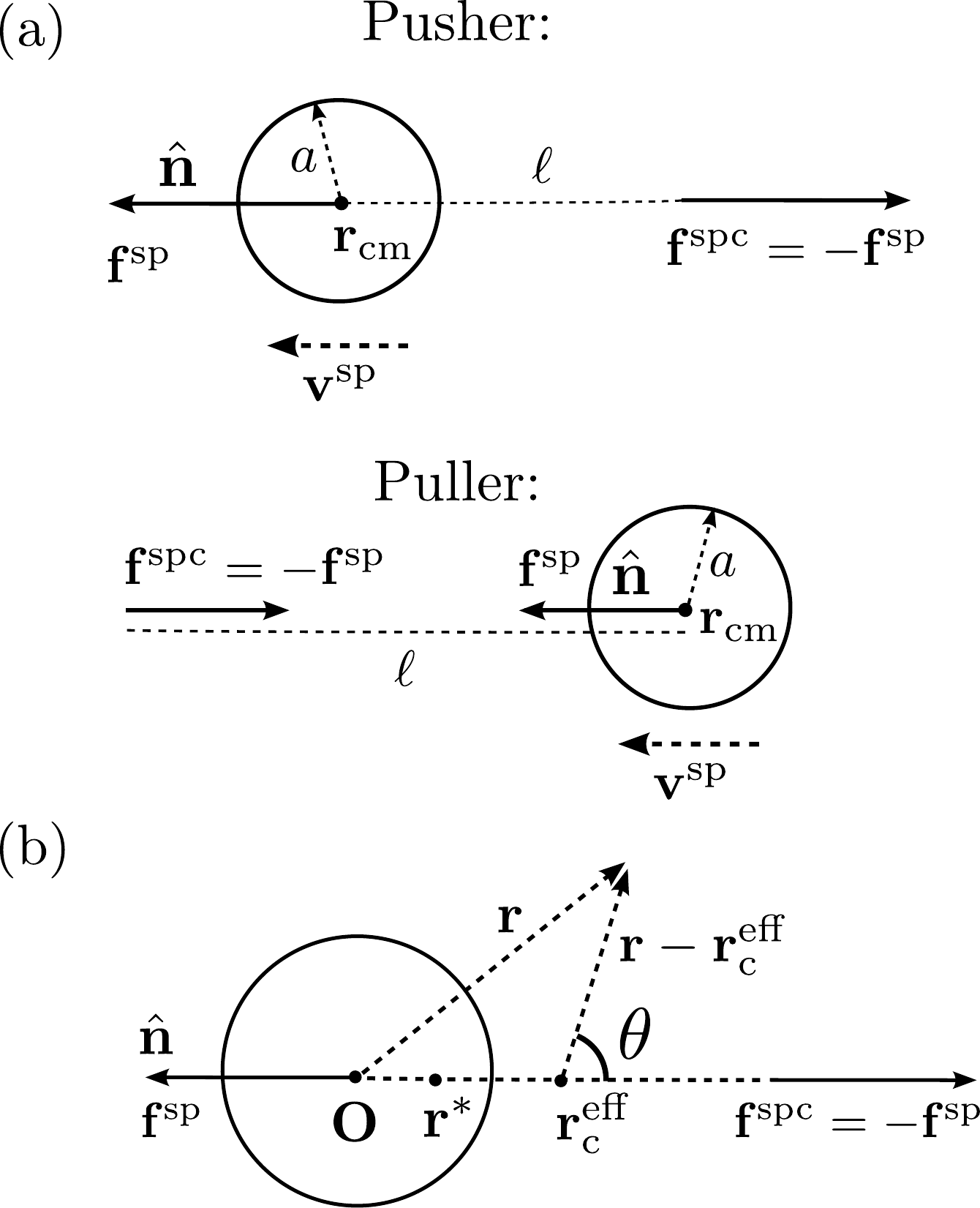}

\caption{(a) Schematics of our minimal model for a spherical swimmer of radius
$a$.  The swimmer is driven by an active force $\fb^{\mathrm{sp}}= f^{\rm{sp}}
\nhat $  
directed along
its intrinsic orientation $\nhat$.  Force-neutrality of the system is
incorporated via a conjugate force $\fb^{\mathrm{spc}} = - \fb^{\mathrm{sp}}$
that acts on the fluid at the position $\rb_{\mathrm{c}}=\rb_{\mathrm{cm}} \pm
\ell\nhat $
\tca{where $\rb_{\mathrm{cm}}$ is the center of mass of the spherical body}.  
For a pusher (top), $\rb_{\mathrm{c}}=\rb_{\mathrm{cm}} -
\ell\nhat $ and the forces $\fb^{\mathrm{sp}}$ and $\fb^{\mathrm{spc}}$ point
away from each other whereas for a puller (bottom),
$\rb_{\mathrm{c}}=\rb_{\mathrm{cm}} + \ell\nhat $, the forces point towards
each other.  (b) The model swimmer in a coordinate system whose origin $\Ob$
coincides with the swimmer's center of mass $\rb_{\mathrm{cm}}$. $\rb$ denotes
any point in the fluid and $\rb_{\mathrm{c}}^{\rm{eff}}$ the center of the
effective point-force dipole that produces the same flow in the far-field as
the swimmer.  Any point in the $x$-$y$ plane containing the vector $\nhat =
-\xhat$ can also be denoted by the vector $\rb - \rb_{\mathrm{c}}^{\rm{eff}}$
which originates from $\rb_{\mathrm{c}}^{\rm{eff}}$ and makes an angle $\theta$
with $\xhat$.  The point $\rb^* = -(a^2/\ell) \nhat$ 
\tca{denotes the position of}
the hydrodynamic image systems of the point-force $\fb^{\mathrm{spc}}$
generated by the sphere.}

\label{schematic_model}
\end{center}
\end{figure}

Our microswimmer model consists of a spherical cell body of radius $a$
self-propelled with a constant force $\fb^{\rm{sp}}$.  The force neutrality
implies that a force with the same magnitude $f^{\rm{sp}}$ and with  the
opposite direction acts on the fluid at a distance $\ell$ from its center of
mass; see Fig. \ref{schematic_model} for a schematics of the model. Although a
similar  model has been proposed before in Ref.
\onlinecite{active_rotor_2012}, its flow field and the hydrodynamic
interactions have been calculated only in  the limit of $a/\ell \to 0$. Here,
we obtain the flow field of such a swimmer for arbitrary values of $a/\ell$
by employing the method of image systems \cite{kim_karilla}. 

Our study shows that the flow field of a rigid-bodied spherical microswimmer
with a finite $a/\ell$  differs from that of a point-force dipole both
qualitatively and quantitatively.  The magnitude and the angular dependence of
the flow field deviate from that of a point-force dipole even at distances
significantly larger than the swimmer's size.  Remarkably, the front-back
symmetry of the flow field  with respect to the self-propulsion direction is
broken.  Eventually, at very large distances  the flow field of this model
converges to that of a point-force dipole.  Nevertheless, its dipole strength
$S^{\mathrm{eff}}$ is different from $f^{\rm{sp}} \ell$, $\it i.e.$, the dipole
strength in the limit $a/\ell \to 0$.  Instead, it is given by
$S^{\mathrm{eff}}=f^{\rm{sp}} \ell^{\mathrm{eff}}$ in which
$\ell^{\mathrm{eff}} < \ell$.

The calculation of the flow field allows us to investigate the influence of the
finite body-size on the hydrodynamic interactions.  For this purpose, we derive
the  \emph{grand} mobility tensor that connects the linear and angular
velocities of the swimmers to the active and passive forces and torques acting
on them. We find that the 
\tca{the leading order}
far field hydrodynamic interactions among the microswimmers 
converge to those among point-force dipoles 
\tca{when}
we renormalize the dipole strengths of the latter to $S^{\rm{eff}}$.
\tca{However, the next leading order terms do not agree even at far distances
unless $a/\ell \to 0$}. 
The explicit expressions for the mobility tensor
allow  us to employ Stokesian dynamics simulations \cite{SD_brady_bossis} for
exploring the collective behavior of rigid-bodied microswimmers at
experimentally relevant concentrations.

\tca{ Before we proceed, we point out that there are other studies besides Ref.
\onlinecite{active_rotor_2012} mentioned above, that have investigated the flow
fields and hydrodynamic interactions of microswimmers with finite bodies
\cite{graham_dumbbell_PRL2005, gyrya_2010, goldstein_PRL2010, Guasto_PRL2010,
dunstan2012, menzel2016, mueller_2017, graaf_JCP, pimponi_2016}.  Menzel {\it
et al.} \cite{menzel2016} have proposed a spherical swimmer model with two
equal and opposite point-forces that act directly on the fluid where the
swimmer's body is placed asymmetrically between the two forces. They also have
disentangled the contributions of active and passive forces to the mobility
tensor.  However, they have  neglected the hydrodynamic contributions of the
image systems of the point forces near the  swimmer's body  in their
calculations. Furthermore, similar swimmer models  with finite size and
elongated body   driven by two point forces have been studied  numerically
using the Lattice-Boltzmann simulation method by Graaf {\it et al.}
\cite{graaf_JCP}.}

The remainder of paper is organized as follows. In section II, we introduce the
minimal swimmer  model and examine its dynamics and flow field in detail.  The
section III is devoted to the hydrodynamic interactions between different
swimmers. Finally, we conclude our work in section IV where  we summarize our
main findings and discuss the consequences of the finite body size on the
dynamics of the swimmers.

\section{ Minimal model for a spherical swimmer and its dynamics}
\label{sect_model}

We first introduce our model for the simplest realization of
rigid-bodied spherical microswimmers.  Then, we analyze the flow properties  of
an individual swimmer both in the absence and the presence of external forces
and torques.

\subsection{Model description}

Our minimal microswimmer model consists of a rigid spherical body with
hydrodynamic radius $a$ suspended in a fluid of viscosity $\eta$.  The swimmer
has an intrinsic orientation $\nhat$ defined by its self-propulsion direction.
We assume  that  no active torque acts on the swimmer and the self propulsion
is provided by an internal force $\fb^{\mathrm{sp}} = f^{\mathrm{sp}} \nhat$.
\tca{As a working rule of the rigid body dynamics, $\fb^{\mathrm{sp}}$ can be
considered to be acting on any arbitrary point along $\nhat$, specially on the
center of mass $\rb_{\rm{cm}}$. However, in Sect. \ref{sect:farfield} we show
that the effective point of action of $\fb^{\mathrm{sp}}$ is actually slightly
shifted from $\rb_{\rm{cm}}$.}

There are a multitude of mechanisms for the generation of self-propulsion
adopted by various microswimmers.  In all the cases there is an internal cycle
of period $\tau$ that repeats itself to generate the propulsion. The force
$\fb^{\mathrm{sp}}$ then represents the thrust experienced by a microswimmer
when averaged over time $t\gg\tau$.  On such time scales  the force-free
condition for a swimmer is satisfied by introducing a point-force
$\fb^{\mathrm{spc}} =- \fb^{\mathrm{sp}}$ acting on the fluid at the position
$\rb_{\mathrm{cm}} \pm \ell \nhat$ where  $ \ell > a$.  The plus sign describes
a puller-type swimmer while the minus sign corresponds to a pusher type. A
schematics of our model swimmer is presented in Fig.  \ref{schematic_model}
(a).

Relating to a real microswimmer, in the case of a flagellated bacterium, for
example, $\fb^{\mathrm{spc}}$ can be interpreted as  the time averaged force
exerted by its flagella on the fluid and $\fb^{\mathrm{sp}}$ is the resulting
thrust force on the body of the bacterium.  For real microorganisms,
$\fb^{\mathrm{spc}}$ is distributed over an extended region in the fluid.  Our
model treats these forces  minimally as an effective  hydrodynamic force acting
at a single point on  the fluid that lies at a distance $\ell$ from
$\rb_{\mathrm{cm}}$ and neglects the details of the force distribution to a
first order approximation.  The distance $\ell$ remains as a free parameter in
the model that can be interpreted as the distance from the center of mass of
the body to an effective action point of the hydrodynamic forces on the fluid.

Due to their small sizes and low speeds, microswimmers move  in the realm of
low Reynolds numbers. Hence, we can neglect  inertial effects both on the
dynamics of the fluid and that of the microswimmers \cite{purcell1977life}.  In
this limit,  the flow field  is described by  the incompressible Stokes
equation and the microswimmer's dynamics is controlled by force and torque
balance equations given by
\begin{eqnarray}
\label{force_balance}
\Fb^h + \fb^{\mathrm{sp}} + \fb^{\mathrm{ext}} &=& 0, \\ 
\label{torque_balance}
\Lb^h + \Lb^{\mathrm{ext}} &=& 0, 
\end{eqnarray}
where   $\Fb^h$ and $\Lb^h$ are the total force and torque exerted by the
surrounding fluid on the swimmer.  $\fb^{\mathrm{ext}}$ and
$\Lb^{\mathrm{ext}}$ are respectively the net external force and torque
experienced by it.   In the following, we   analyze the motion of a 
swimmer in the absence and presence of external forces and torques. 

\subsection{Dynamics of a free swimmer}

We first focus on the dynamics of a free microswimmer {\it i.e.} one on which
no external forces or torques are exerted.  We  obtain the self-propulsion
velocity in terms of the model parameters.  Then, we discuss the microswimmer's
flow field and its far-field behavior. 

\subsubsection{Self-propulsion velocity}

\begin{figure}
\begin{center}
\includegraphics[width=.95\columnwidth]{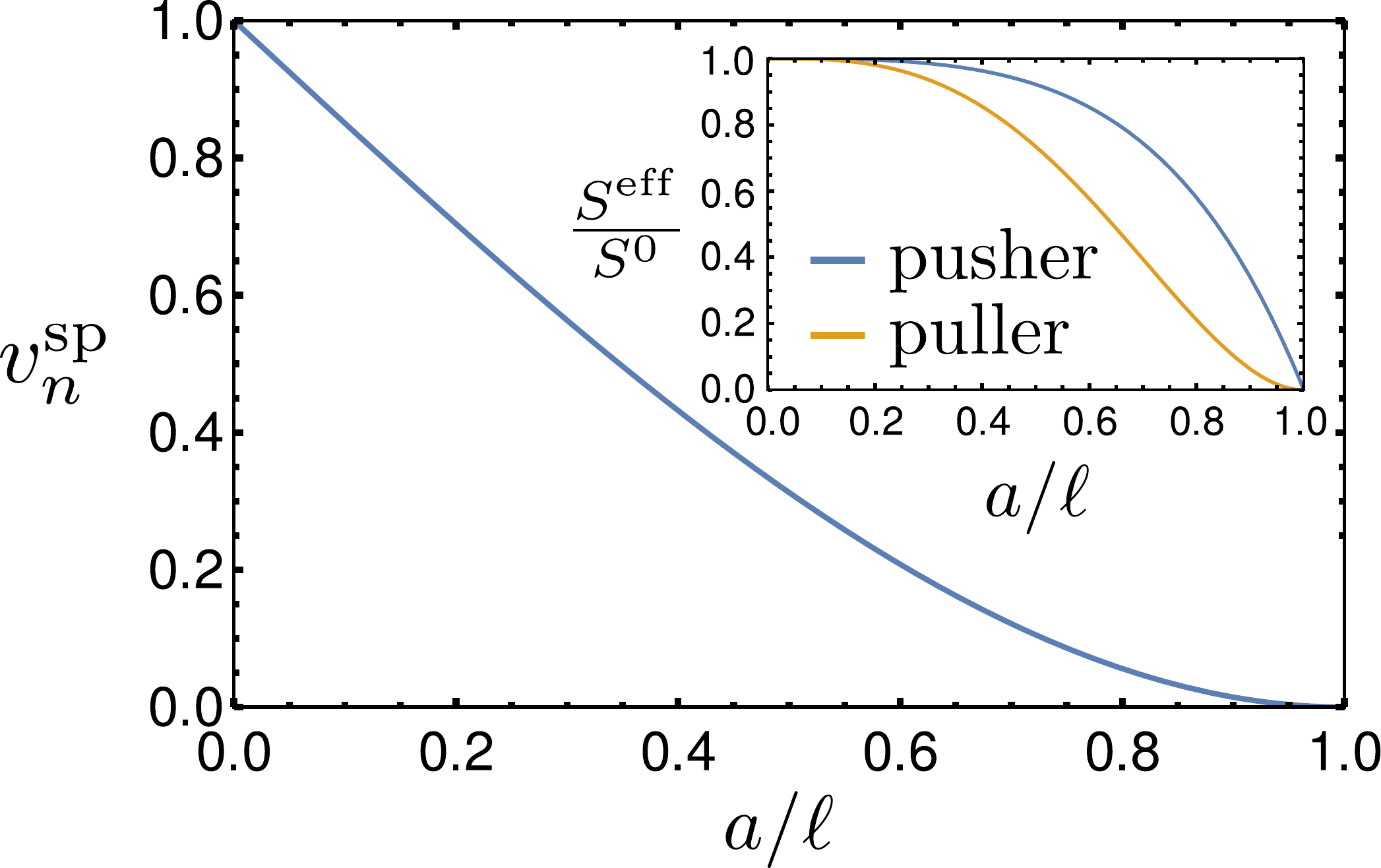}

\caption{The normalized swimming speed $v^{\mathrm{sp}}_n =
v^{\mathrm{sp}}/(f^{\mathrm{sp}}/6\pi\eta a)$ of a spherical swimmer of radius
$a$, plotted against $a/\ell$, where $\ell$ is the separation between the
active forces $\fb^{\mathrm{sp}}$ and $\fb^{\mathrm{spc}} = -
\fb^{\mathrm{sp}}$. The plots of $v^{\mathrm{sp}}_n$  for both pushers and
pullers are identical. Inset: The ratio $S^{\mathrm{eff}}/S^0$ of the dipole
strengths of the {\em effective} point-force dipole producing the correct
far-field and the point-force dipole obtained in the limit of zero
swimmer-size.}

\label{v_swim_plot}
\end{center}
\end{figure}

In order to obtain the net self-propulsion velocity of the free swimmer
$\vb^{\mathrm{sp}}$, we assume that the fluid velocity on the surface of the
spherical body satisfies the stick boundary conditions.  Hence, the velocity of
any point $\rb_S$ on its surface $S$ is equal to the local fluid velocity at
that point:
\begin{eqnarray}
\vb^{\mathrm{sp}}
    &=& - \oint_S dS^{\prime}\, \Tb (\rb_S - \rb_S^{\prime})\cdot \fb(\rb_S^{\prime}) \nonumber \\
    && + \, \ub_0 (\rb_S),          
\label{v_swim}
\end{eqnarray}
where  $\Tb(\rb) = (1/8\pi\eta r) (\Ib + \rb\rb/r^2)$ is the Oseen tensor, and
$\ub_0 (\rb) = - \Tb(\rb - \rb_{\mathrm{cm}} \pm \ell \nhat)\cdot
\fb^{\mathrm{sp}}$ is the flow field because of the point-force
$\fb^{\mathrm{spc}}$ acting on the fluid.  Here   $+ $ and $-$  denote  a
pusher and puller type of swimmer respectively.  $\fb(\rb_S)$ is the force per
unit area exerted by the fluid on the swimmer's surface at $\rb_S$.  Note that
$\fb(\rb_S)$ acquires a contribution from the fluid flow created by the
point-force $\fb^{\mathrm{spc}}$. Thus, it is not uniform over the spherical
surface as it would have been in the absence of $\fb^{\mathrm{spc}}$
\cite{dhont}.  Furthermore, the angular velocity of the swimmer does not appear
in Eq. (3) because no external torque acts on the body and the flow $\ub_0$
does not generate any rotation of the swimmer \cite{zeroangularfoot}.

Integrating both sides of Eq. (\ref{v_swim}) over the surface of the swimmer
and utilizing the spherical symmetry of the body \cite{integration_oseen_foot},
we obtain
\begin{eqnarray}
\vb^{\mathrm{sp}} = - \frac{1}{6 \pi \eta a} \oint_S dS^{\prime}\, \fb(\rb_S^{\prime}) 
      + \frac{1}{4\pi a^2} \oint_S dS\, \ub_0(\rb_S),
\label{pre_faxen}
\end{eqnarray}
where $\oint_S dS^{\prime}\,\fb(\rb_S^{\prime})$ is equal to the net
hydrodynamic force $\Fb^{\mathrm{h}}$ on the sphere.  The force balance
condition, Eq. (\ref{force_balance}), for a free swimmer  thus implies
that  $\oint_S dS^{\prime}\,\fb(\rb_S^{\prime}) = - \fb^{\mathrm{sp}}$.  The
second integral above can be calculated by a Taylor expansion of the flow field
around $\rb_{\mathrm{cm}}$ \cite{dhont}. 
\tca{Because of the spherical symmetry of the swimmer the Taylor
expansion truncates at the quadratic order in $a$. Thus, Eq. (\ref{pre_faxen})
simplifies to the exact expression,}
\begin{eqnarray}
\vb^{\mathrm{sp}} =   \frac{1}{6\pi\eta a} \fb^{\mathrm{sp}} 
      + \left(1 + \frac{a^2}{6} \nabla^2_{r_{\mathrm{cm}}} \right) \ub_0(\rb_{\mathrm{cm}}).
\label{faxen}
\end{eqnarray}
This equation is identical to the familiar Faxen's law \cite{dhont}.
Evaluating  $\nabla^2_{r_{\mathrm{cm}}}  \ub_0(\rb_{\mathrm{cm}})$ explicitly,
we obtain the self-propulsion velocity in terms of $a$ and $\ell$:
\begin{eqnarray}
\vb^{\mathrm{sp}} =   \frac{1}{6\pi\eta a} \left\{1 - g_{a/\ell}\right\}\fb^{\mathrm{sp}}, 
\label{v_swim_final}
\end{eqnarray}
in which $g_{a/\ell} \equiv (3/2)[a/\ell - (a/\ell)^3/3]$.  Since, $0 < g \le
1$ for the physically relevant range $0 < a/ \ell\le 1$, the swimmer's velocity
is in the direction of $\fb^{\mathrm{sp}}$, as expected.  Furthermore, owing to
the symmetry of the Oseen tensor  $\Tb(+\ell \nhat) = \Tb(-\ell \nhat)$, for  a
given set of parameters $(a,\ell)$, the self-propulsion speeds of a pusher and
a puller are identical.

In Fig. \ref{v_swim}, we have presented  the normalized swimming speed $
v^{\mathrm{sp}}_n = v^{\mathrm{sp}}/(f^{\mathrm{sp}}/6\pi\eta a)$ versus
$a/\ell$.  We note  that $v^{\mathrm{sp}}_n$ is a decreasing function of
$a/\ell$.  In the limit  of $a\ll\ell$, the speed of the swimmer approaches to
that of a sphere dragged by a single isolated force $f^{\mathrm{sp}}$,  {\it
i.e.} $v^{\mathrm{sp}}\rightarrow f^{\mathrm{sp}}/6\pi\eta a$.  However, for
$a/\ell \ge 0.4$,  a range relevant for some of flagellated bacteria,    the $
v^{\mathrm{sp}}_n $ value is  considerably smaller than one.   Thus the flow
created by the neutralizing force $\fb^{\mathrm{spc}}$ results in a
considerable lowering of the speed of the swimmer compared to its passive
counterpart -- an externally driven sphere.

\subsubsection{Flow field}

Next, we obtain the flow field of a spherical swimmer and  compare it to that
of a point-force dipole.  To calculate the flow field $\ub(\rb)$ of the
swimmer,  we need to incorporate two contributions: (a) the flow field $\ub_a$
generated because of  the point-force $\fb^{\mathrm{spc}}$ that lies near a
sphere and (b) the flow $\ub_b$ resulted from the translation of the sphere
with velocity  $\vb^{\mathrm{sp}}$ in an otherwise quiescent flow.  In the
following, we obtain each of these contributions for a sphere  centered at the
origin $\Ob$.

$\ub_a$ can be expressed as a superposition of the flow created directly by the
point-force and the modifications to it, denoted as $\ub^*(\rb)$,  due to the
surface of the sphere:
\begin{eqnarray}
\ub_a(\rb) = - \Tb(\rb \pm \ell \nhat)\cdot \fb^{\mathrm{sp}} + \ub^*(\rb).    
\label{flow_fspc}
\end{eqnarray}    
The field $\ub^*(\rb)$ can effectively be ascribed to an image system of the
point-force created by the sphere \cite{sphere_pt_force}.  It  is given by
\tca{the exact expression,}
\begin{eqnarray}
\ub^*(\rb) &=& g_{a/\ell}\, \fb^{\mathrm{sp}} \cdot \Tb(\rb - \rb^*) \nonumber \\ 
           && -\, h_{a/\ell}\, a f^{\mathrm{sp}} 
                 \left[\left(\nhat\nhat - {1\over 3}\Ib\right)\cdot\nablab\right]\cdot\Tb(\rb - \rb^*) \nonumber \\
           && +\, j_{a/\ell}\, a^2 \fb^{\mathrm{sp}} \cdot \nabla^2 \Tb(\rb - \rb^*), 
\label{image_field}
\end{eqnarray}
where $g_{a/\ell} = (3/2)[a/\ell - (a/\ell)^3/3]$ as defined earlier,
$h_{a/\ell}= a^2/\ell^2 -a^4/\ell^4$ and $j_{a/\ell}= (1/4)(a/\ell)(1
-a^2/\ell^2)^2$. 
\tca{The point $\rb^* = \mp \ell^* \nhat$, with $\ell^* := a^2/\ell$ denotes
the position of the image system of the point-force $\fb^{\mathrm{spc}}$
\cite{sphere_pt_force}, see Fig.  \ref{schematic_model} (b). The singularity solutions for
the image system of the point force  consist of a Stokeslet (point force), a stresslet
and a degenerate quadrupole (source dipole) with strengths $g_{a/\ell}, h_{a/\ell}$ and
$2j_{a/\ell} a^2$, respectively \cite{sphere_pt_force}.} 
We note that $\ub_a(\rb_S) = 0$ at any point $\rb_S$ on the surface of the
swimmer satisfying the no-slip boundary condition for a stationary sphere.

The flow field $\ub_b$ set up by a sphere translating with the velocity
$\vb^{\mathrm{sp}}$ is obtained as \cite{dhont}
\begin{eqnarray}
\ub_b (\rb) = 8\pi\eta \left[{3 a\over 4} \Tb(\rb) 
                    + {a^3\over 8} \nabla^2 \Tb(\rb)\right] \cdot \vb^{\mathrm{sp}},
\label{flow_past_sph}
\end{eqnarray}
where $\vb^{\mathrm{sp}}$ is  given by Eq.  (\ref{v_swim_final}). Consequently,
the total flow field of a spherical swimmer reads as:
\begin{eqnarray}
\ub^{\mathrm{sp}}(\rb) &=& \ub_a (\rb) + \ub_b (\rb) \nonumber \\
         &=& \left\{(1-g_{a/\ell}) \left[\Tb(\rb) - {a^2\over 3} \widetilde{\Tb}(\rb)\right] 
             - \Tb(\rb \pm \ell \nhat)  \right. \nonumber \\
         &&  +\, g_{a/\ell}\, \Tb(\rb \pm \ell^*\nhat) 
            \left. - 2 j_{a/\ell}\, a^2 \widetilde{\Tb}(\rb \pm \ell^*\nhat) 
            \vphantom{{a^2\over 3}}\right\}\cdot \fb^{\mathrm{sp}} \nonumber \\ 
         && +\, h_{a/\ell}\, a f^{\mathrm{sp}} \Hb(\rb \pm \ell^*\nhat, \nhat)\cdot(\rb \pm \ell^*\nhat). 
\label{flow_swimmer}
\end{eqnarray}
Here, the $+$ and $-$ signs apply to the pusher and puller types, respectively.
\tca{For brevity, we have defined the new tensors, $\widetilde{\Tb}(\rb) =-
(1/2) \nabla^2 \Tb(\rb) = (1 \big/ 8\pi\eta r^3) (-\Ib + 3 \rb\rb/r^2)$  and
$\Hb(\rb,\nhat) = (1\big/ 8\pi\eta r^3) [-\Ib + 3 (\nhat\cdot\rb)^2  \Ib/r^2]$.
Note that $\widetilde{\Tb}(\rb) \cdot \nhat$ corresponds to the flow field of a
degenerate quadrupole and $\Hb(\rb,\nhat)\cdot \rb$ gives the flow field of a
force dipole oriented in the direction $\nhat$.  Since Eqs.
\eqref{flow_fspc}-\eqref{flow_past_sph} are exact, Eq.  \eqref{flow_swimmer}
represents the exact form of the flow field for our model swimmer valid at any
distance $r$.}

\begin{figure}
\begin{center}
\includegraphics[width=.8\columnwidth]{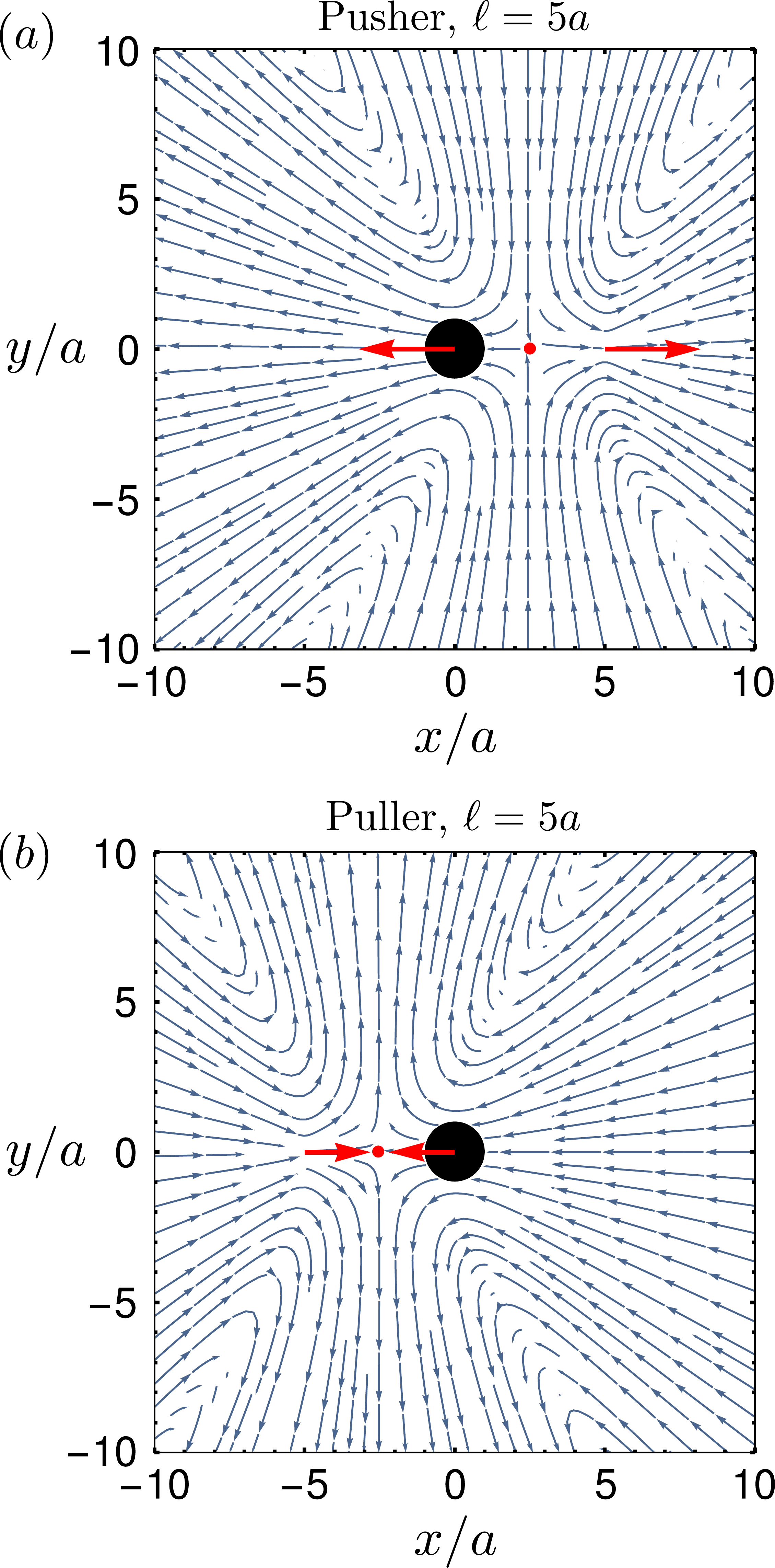}

\caption{The flow streamlines (blue arrows)  created by a spherical
swimmer of radius $a$, with orientation $\nhat = - \xhat$ and $\ell = 5 a$,
plotted in the $x$-$y$ plane. Panels (a) and (b) correspond  to a pusher and
a puller type swimmer, respectively. The red arrows represent the active forces
$\fb^{\mathrm{sp}}$ and $\fb^{\mathrm{spc}} = - \fb^{\mathrm{sp}}$ and the red
dot indicates the effective center $\rb_{\rm{c}}^{\mathrm{eff}}$ of the equivalent far-field
point-force dipole in each case.}

\label{streamlines_push-pull}
\end{center}
\end{figure}

In Fig.  \ref{streamlines_push-pull}, we have plotted the streamlines of the
flow fields  of a pusher and a puller respectively, described by Eq.
(\ref{flow_swimmer}), in a plane containing their orientation vectors $\nhat =
- \xhat$.  As expected, for a pusher (puller), there is an outward (
inward) flow along the axis containing $\nhat$ and an inward (outward) flow
normal to that.

\subsubsection{Multipole expansion of the flow field}
\label{sect:farfield}

Having \tca{obtained} the full flow field of the microswimmer, we now
investigate its far-field behavior. 
\tca{An axisymmetric rotation-free (no active torque) swimmer, with its body
centered at the origin and self-propelling in the direction $\nhat$ generates a
far-field  flow  of the general form  \cite{lauga_fluidmech2012}
\begin{eqnarray}
\label{eq:multipole}
\ub^{\mathrm{sp}}_{r \gg a} (\rb) &=& S^{\mathrm{eff}} (\hat{\nb} \cdot \nablab)[\Tb(\rb)\cdot \hat{\nb}] \\ \nonumber
    &-&\frac{1}{2}D^{\mathrm{eff}} \nabla^2 [\Tb(\rb)\cdot \hat{\nb}] \\ \nonumber
    &+& Q^{\mathrm{eff}} (\hat{\nb} \cdot \nablab)^2  [\Tb(\rb)\cdot \hat{\nb}]+ O(1/r^4).
\end{eqnarray}
These terms  describe the  leading order  singularity solutions of the Stokes
equation for a force-free swimmer. They correspond to a force dipole, a
degenerate quadrupole (source dipole)  and a force quadrupole, respectively.
The  coefficients  $S^{\mathrm{eff}}$, $D^{\mathrm{eff}}$ and
$Q^{\mathrm{eff}}$  characterize the strengths of the respective multipoles and
their values depend  on the swimmer model under consideration.  The flow filed
of the force dipole  decays as $1/r^2$ and those of the two quadrupolar terms
as $1/r^3$. To  obtain the multipolar strengths for our model, we expand
$\ub^{\mathrm{sp}}(\rb)$ given in Eq. (\ref{flow_swimmer}) in powers of the
inverse distance $1/r$ from the center of mass of the swimmer.}

\tca{We find that for our model the first leading order term at far distances
$(r \gg \ell > a$)  decays as  $1/r^2$ and  is given by}
\begin{eqnarray}
\label{eq:dipole}
\ub^{\mathrm{sp}}_{r \gg \ell} (\rb) = \pm \frac{1}{8\pi\eta} \frac{S^{\mathrm{eff}}}{r^2}
                      \left[-1 + 3 (\rhat\cdot\nhat)^2\right]\rhat. 
\end{eqnarray}
It  is identical to the flow field of a point-force dipole  with an effective dipole strength
\begin{equation}
\label{eq:dipoleS}
\tca{S^{\mathrm{eff}} = S^0[1- (a/\ell)^2g_{a/\ell} \pm (a/\ell) h_{a/\ell}],} 
\end{equation}
\tca{where $+$ and $-$ refer to a pusher and puller respectively.}
Here, $S^0 = \ell f^{\mathrm{sp}}$ represents the  dipole strength of a
point-force dipole consisting of two point-forces of magnitude
$f^{\mathrm{sp}}$ that lie at a distance $\ell$ apart.   $S^0$ is \tca{identical to the} {\em
effective} dipole strength of the swimmer when $a/\ell\to 0$.  However, for an
accurate far-field representation of a finite sized swimmer by a point-force
dipole, the driving forces $f^{\mathrm{sp}}$ must be separated by an effective
distance 
\tca{$\ell^{\mathrm{eff}} = \ell [1- (a/\ell)^2 g_{a/\ell} \pm (a/\ell)
h_{a/\ell}]< \ell$.  This implies that the effective center of the force on the
fluid by the surface of the swimmer, $-\oint_S dS^{\prime}\,
\fb(\rb_S^{\prime})$, or equivalently, the point of action of the force
$\fb^{\mathrm{sp}}$ on the swimmer, is situated at $-\nhat (\ell -
\ell^{\rm{eff}})$, which is slightly shifted from the center of mass
$\rb_{\rm{cm}}$ of the sphere. For clarity,   {\it associated point-force
dipole} is defined as a point-force dipole  that consists of  two point-forces
$\pm \fb^{\rm{sp}}$  separated by a distance $\ell^{\rm{eff}}$. The center of
the {\it associated point-force dipole} lies at $\rb_{\rm{c}}^{\mathrm{eff}} =
-\nhat (\ell
- \ell^{\rm{eff}}/2)$ [See Fig. \ref{streamlines_push-pull}].  Similarly, by a
  {\it bare} point-force dipole we refer to the dipole obtained in the limit of
$a/\ell \to 0$, where the two point-forces are separated by a distance $\ell$.}

\tca{In the inset of Fig. \ref{v_swim_plot}, we have shown $S^{\mathrm{eff}}$
as a function of $a/\ell$ for both a pusher and a puller. We see that the
effective dipole strength is renormalized differently for pushers and pullers.
This difference results from the symmetry of the image stresslet  (the term $
\propto h_{a/\ell}$ in Eq. \eqref{image_field}) which always acts like a
pusher. Thus, this term enhances the dipolar strength of a pusher, whereas it
reduces that of a puller.} Evidently, when $\ell \gg  a$, the swimmer's size
can be neglected and $S^{\mathrm{eff}} \to S^0$.  Otherwise,  even for the
far-field flow, the finite size of the swimmer should be taken into account  by
renormalizing  the dipole strength to $S^{\mathrm{eff}}$.

\begin{figure*}
\begin{center}
\includegraphics[width=0.98\linewidth]{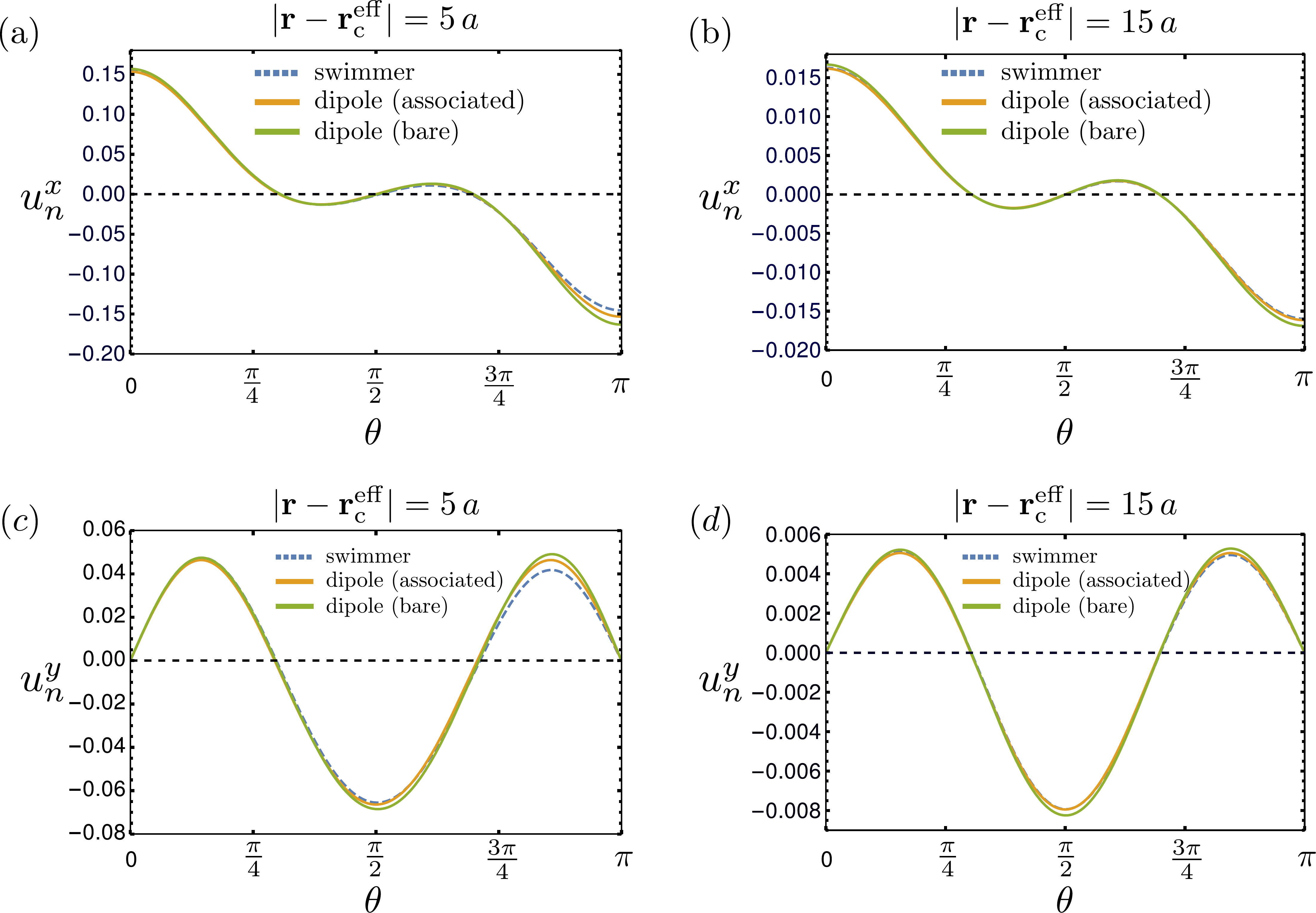}

\caption{ \tca{Angular dependence of the flow fields of the swimmer and its
associated point-force dipole and also that of the bare point-force dipole
obtained in the limit of zero swimmer size. For each case, the normalized flow
field $u_{\rm{n}}^{\alpha}(|\rb - \rb_{\rm{c}}^{\rm{eff}}|, \theta) =
u^{\alpha} (|\rb - \rb_{\rm{c}}^{\rm{eff}}|, \theta)/(f^{\rm{sp}}/6\pi\eta a)$
with $\alpha = x,y$, is shown as a function of the angle $\theta$ about the
effective center $\rb_{\rm{c}}^{\mathrm{eff}}$ measured at distances $|\rb
- \rb_{\rm{c}}^{\rm{eff}}| = 5\, a$ [(a) and (c)] and $15\,a$. [(b) and (d)].} }

\label{bare_Vs_assoc}
\end{center}
\end{figure*}

\tca{We depict the effect of such a renormalization on the flow field in Fig.
\ref{bare_Vs_assoc}. The figure shows the angular dependence of the  normalized
components of the flow-field of a swimmer moving in $-\hat{\xb}$ direction,
{\it i.e.} $u_{\rm{n}}^{\alpha} (|\rb -\rb_{\rm{c}}^{\rm{eff}}|, \theta) =
u^{\alpha}(|\rb - \rb_{\rm{c}}^{\rm{eff}}|, \theta)/(f^{\rm{sp}}/6\pi\eta a)$
where $\alpha=x,y$.  For comparison, we have also included the corresponding
flow fields of the associated and bare point-force dipoles.  Here, $|\rb
-\rb_{\rm{c}}^{\rm{eff}}|$ is the distance measured from the effective center
$\rb_{\rm{c}}^{\mathrm{eff}}$ and $\theta$ is the angle between $\rb
-\rb_{\rm{c}}^{\rm{eff}}$ and $\xhat = - \nhat$ [Fig.  \ref{schematic_model}
(b)].  We find that at short distances [$r=5a$ in Fig.  \ref{bare_Vs_assoc} (a)
and (c)] there is a considerable difference between the flow fields of the
swimmer and the bare point-force dipole, specially near $\theta=\pi$. However,
the difference is reduced for the associated point-force dipole but still
visible for the  directions in front of the swimmer, especially for $\theta > 3
\pi/4$.  At farther distances, {\it e.g.}, at $r=15a$,  as presented in Fig.
\ref{bare_Vs_assoc} (b) and (d), we find a very good agreement between the flow
fields of the swimmer and its associated point-force dipole but they
considerably differ from that of of the bare point-force dipole. }

\tca{Subsequently, we obtain the coefficients of the next leading order terms
($\propto 1/r^3$).  The strength of the degenerate quadrupolar  term
$\frac{1}{2} \hat{\nb} \cdot \nabla^2 \Tb(\rb)$ is given by
\begin{equation}
 D^{\mathrm{eff}} = -[2j_{a/\ell}+ (1-g_{a/\ell})/3] f^{\mathrm{sp}} a^2.
\end{equation} 
$D^{\mathrm{eff}}$  is a function of both  $a$ and $\ell$ as it stems from the
finite size of the cell body.  The strength of the force quadrupolar term 
$(\hat{\nb} \cdot \nabla)^2 \Tb(\rb)$ is given by
\begin{equation}
 Q^{\mathrm{eff}}= - \left[ 1- g_{a/\ell} (a/\ell)^4 \pm (a/\ell)^3 h_{a/\ell} \right] f^{\mathrm{sp}} \ell^2.
\end{equation} 
$Q^{\mathrm{eff}}$  results  from the length asymmetry between size of the cell
body $a$ and the effective length of the flagellum $\ell$ \cite{lauga_fluidmech2012}.}
\tca{Note that unlike the dipolar strength, the quadrupolar strengths have the
dimension of force times length squared and  they vanish for a point-like
swimmer, {\it i.e.} $a \rightarrow 0$ and $\ell \rightarrow 0$.}

\subsubsection{Swimmer's finite size effects on  the flow field }
\label{size_eff_flow}

\begin{figure}
\begin{center}
\includegraphics[width=0.99\columnwidth]{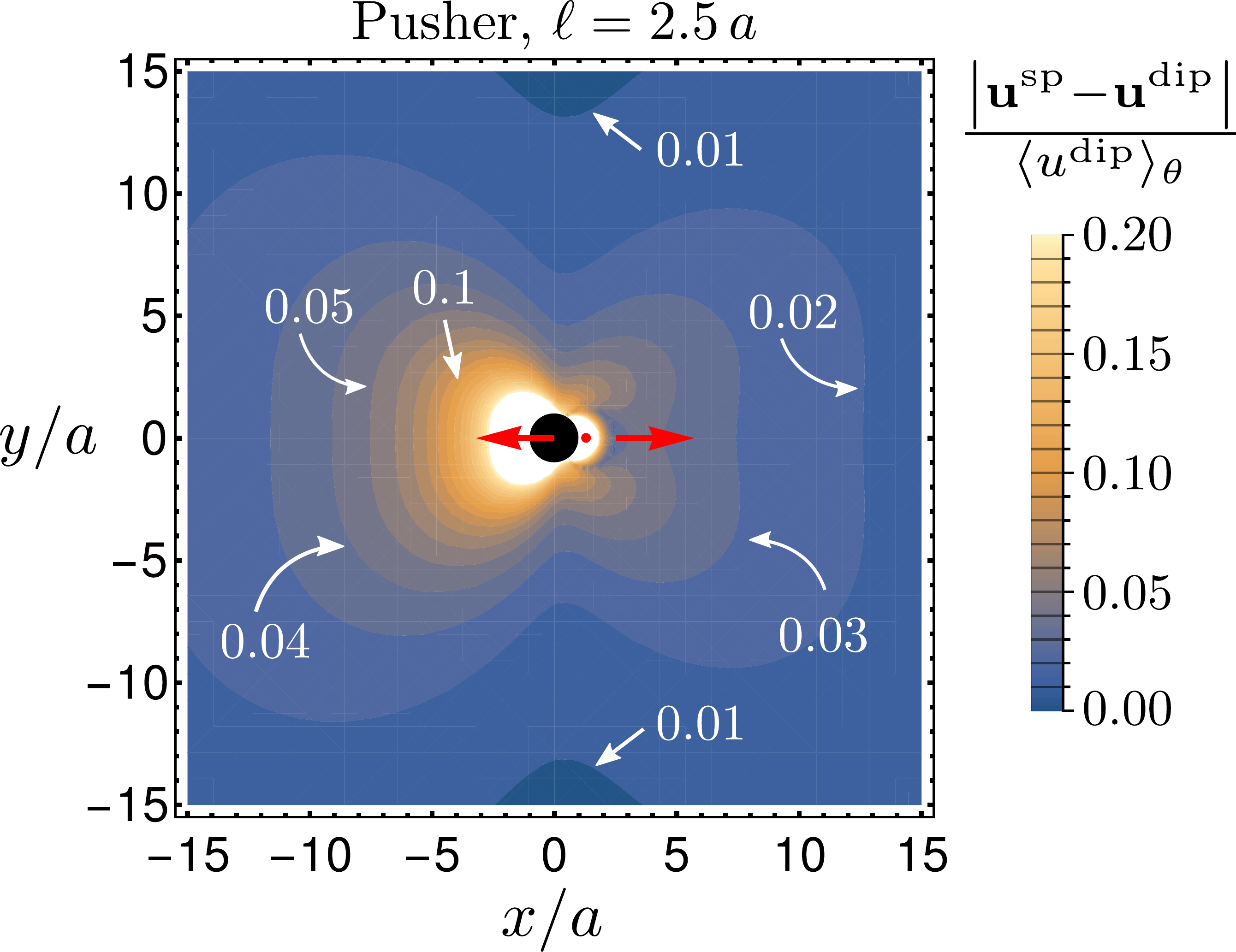}

\caption{ \tca{Comparision of the flow fields of the swimmer with $\ell = 2.5
a$ (pusher) and its associated point-force dipole. The figure shows the contour
plot of the magnitude of the difference of the flows of the swimmer and its
associated point-force dipole, $\left|\ub^{\rm{sp}}(\rb) -
\ub^{\rm{dip}}(\rb)\right|$, normalized by the angular averaged flow speed of
the dipole, $\langle u^{\rm{dip}} \left(\left|\rb -
\rb_{\rm{c}}^{\rm{eff}}\right|\right)\rangle_\theta$ at the corresponding
distance from the center $\rb_{\rm{c}}^{\rm{eff}}$. The body of the swimmer is
shown in black; the active forces $\fb^{\mathrm{sp}}$ and $\fb^{\mathrm{spc}}$
and the effective center $\rb_{\rm{c}}^{\mathrm{eff}}$ of the associated
point-force dipole are shown in red. The white region (lighter color)
represents values in the range $0.2$-$0.8$ not shown in the color bar.}}

\label{u-by-udip}
\end{center}
\end{figure}

Next, we  examine the flow field at intermediate distances 
\tca{more closely}
and compare it to that of a point-force dipole. For simplicity we focus on the
$x$-$y$ plane.  In the following, $\ub^{\mathrm{sp}}$ and $\ub^{\mathrm{dip}}$
denote the swimmer's flow field  and  that of \tca{its associated}
point-force dipole with the strength $S^{\mathrm{eff}}$ respectively. In Fig.
\ref{u-by-udip}, we have presented the contour plot of 
\tca{the difference of the two flow fields, $\left|\ub^{\mathrm{sp}}(\rb)
-\ub^{\mathrm{dip}}(\rb)\right|$ normalized by
the angular averaged flow speed of the dipole, $\langle u^{\rm{dip}}
\left(\left|\rb - \rb_{\rm{c}}^{\rm{eff}}\right|\right)\rangle_\theta$, 
for a pusher with $\ell = 2.5 a$. }

\tca{ From Fig. \ref{u-by-udip}  we notice that the angular dependence of the
swimmer's flow field at intermediate distances is singnificantly different from
that of a point-force dipole. In particular, the front-back symmetry of the
flow field of a point-force dipole is broken by the finite size of the
swimmer's body.  This dissimilaity is visible in the head-tail asymmetry  of
the shapes of the contours  around the swimmer.  For example, the contour for
$0.05$ ($5\%$ difference in the flow speeds) extends up to a distance of $\sim
8 a$ from $\rb_{\rm{c}}^{\rm{eff}}$ towards the head, whereas, at similar
distances towards the back the normalized difference is $< 0.03$ ({\it i.e.},
$< 3$\%). Furthermore, contours of any given normalized difference encircle a
larger area in the front than in the back. Hence, the flow field of the
swimmer matches that of the dipole at a shorter distance in any general
direction behind the swimmer ({\it i.e.}, for $\theta < \pi/2$ ) while at the
same distance in a corresponding direction in the front ({\it i.e.}, for
$\theta + \pi/2$ ), we notice remarkable differences in agreement with our
observations from Fig. \ref{bare_Vs_assoc}.}

\tca{The differences between the flow field of the swimmer and that of its
associated force dipole with the same strength reflect the importance of the
next leading order singularity contributions (quadrupolar, octoupolar, etc.)
to the flow field at short and intermediate distances. These higher order
singularity contributions are important for the collective behavior of the
microwswimmers and need to be accounted for in the many-body simulations.  }

\subsection{Dynamics of a swimmer exposed to external forces and torques
\label{sec:ext_f_L}}

An external force acting on a swimmer supplements to its self-propulsion
velocity $\vb^{\mathrm{sp}}$.  Similarly, an external torque
$\Lb^{\mathrm{ext}}$ causes the swimmer to rotate.  Note that the flow created
by self-propulsion of the swimmer does not  generate any angular velocity in
the absence of other swimmers \cite{zeroangularfoot}.  Thus, the translational
and angular  velocities of the swimmer in the presence of the external force
$\fb^{\mathrm{ext}}$ and torque $\Lb^{\mathrm{ext}}$ are modified as
\begin{eqnarray} 
\label{vel_with_ext}
\vb     &=&  \vb^{\mathrm{sp}} + \frac{1}{6\pi\eta a} \fb^{\mathrm{ext}}, \\
\omegab &=& \frac{1}{8\pi\eta a^3} \Lb^{\mathrm{ext}}.
\label{omega_with_ext}
\end{eqnarray}

The flow-field $\ub_b(\rb)$ resulting from the translation of the sphere is
accordingly altered. The contribution of  the external force to the flow  can
be obtained  by substituting $\vb^{\mathrm{sp}}$ with $ \frac{1}{6\pi\eta a}
\fb^{\mathrm{ext}}$ in Eq. (\ref{flow_past_sph}).  Likewise, the rotation of
the swimmer  generates a flow field  which is described by $(a^3/r^3) \omegab \times \rb$
\cite{dhont}. Given the linearity of Stokes equation, the  net flow field of
the swimmer in the presence of   external forces and torques  is thus given by: 
\begin{eqnarray}
\ub(\rb) &=& \ub^{\mathrm{sp}}(\rb) + \left[\Tb(\rb) + {a^2\over 6} \nabla^2 \Tb(\rb)\right]
                        \cdot\fb^{\mathrm{ext}} \nonumber \\ 
         &&   + \frac{1}{8\pi\eta r^3} \Lb^{\mathrm{ext}} \times \rb. 
\label{flow_individual}
\end{eqnarray}
in which the $\ub^{\mathrm{sp}}(\rb)$ is obtained from Eq. \ref{flow_swimmer}.

Having discussed the motion of a single swimmer, we devote the next section to
the hydrodynamic interactions among an ensemble of swimmers in a suspension.

\section{Hydrodynamic interactions among spherical swimmers}

In a suspension of $N$ swimmers, the flow field generated by  each swimmer will
affect the motion of all the others.  In this section, we obtain the grand
mobility tensor that connects the translational and angular velocities of the
swimmers to the forces and torques acting on them. Especially, we elucidate the
contribution of the active forces to hydrodynamic interactions.

The  translational  velocity  $\vb_i$ and angular velocity $\omegab_i$  of the
$i$th swimmer are coupled to the forces  and torques acting on all the swimmers
via the grand mobility tensors defined as below:
\begin{eqnarray}
\left[\begin{array}{c} 
  \vb_i \\ 
  \omegab_i 
\end{array}\right] &=& \sum_{j=1}^N \left[\begin{array}{cc} 
                                            \Mb_{ij}^{\mathrm{tt}} & \Mb_{ij}^{\mathrm{tr}}  \\ 
                                            \Mb_{ij}^{\mathrm{rt}} & \Mb_{ij}^{\mathrm{rr}}  
                                          \end{array}\right] 
                                    \left[\begin{array}{c} 
                                            \fb_j^{\mathrm{ext}} \\ 
                                            \Lb_j^{\mathrm{ext}} 
                                          \end{array}\right] \nonumber \\ 
                 &&  + \sum_{j=1}^N \left[\begin{array}{cc} 
                                            \mutens_{ij}^{\mathrm{tt}} & \mutens_{ij}^{\mathrm{tr}}  \\
                                            \mutens_{ij}^{\mathrm{rt}} & \mutens_{ij}^{\mathrm{rr}}  
                                        \end{array}\right]
                                    \left[\begin{array}{c} 
                                            \fb_j^{\mathrm{sp}} \\ 
                                            \mathbf{0} 
                                          \end{array}\right].
\label{full_mob_matrix}
\end{eqnarray}
Here, the subscripts $i,j$ are the swimmer indices running from $1$ to $N$, and
the superscripts $\mathrm{t}$ and $\mathrm{r}$, stand for `translational' and
`rotational' degrees of freedom, respectively.  The  mobility tensors
$\Mb_{ij}^{\alpha \beta}$ couple the $i$th swimmer's translational and angular
velocities (specified by the index $\alpha$)   to the passive external forces
$\fb_j^{\mathrm{ext}}$ and torques $\Lb_j^{\mathrm{ext}}$ (distinguished by the
index $\beta$) acting on the $j$th swimmer. Accordingly, the tensors $\Mb$ are
identical to the corresponding mobilities of passive spheres.  In contrast, the
active mobility tensors $\mutens$ incorporate the hydrodynamic interactions resulting
from the active forces.  They couple the velocities of the $i$th swimmer to the
active forces $\pm \fb_j^{\mathrm{sp}}$ that propel  the $j$th swimmer.  Below,
we  calculate the explicit  forms of the $\mutens$ tensors  for this model
system.

 \subsection{Self-mobility tensors}

We first focus on the self-mobility tensors.  $\Mb_{ii}^{\mathrm{tt}}$ and
$\Mb_{ii}^{\mathrm{rr}}$ represent respectively  the  translational and
rotational self-mobility tensors of the $i$th sphere  in the absence of
self-propulsion   \cite{dhont}. $\mutens_{ii}^{\mathrm{tt}}$  and
$\mutens_{ii}^{\mathrm{rr}}$ denote the corresponding active self-mobility
tensors.  To simplify the notation, in what follows,  we introduce the
following abbreviations:
\begin{eqnarray}
M^{\mathrm{t}} = {1\over 6\pi\eta a}, \hspace{0.4cm} M^{\mathrm{r}} = {1\over 8\pi\eta a^3}
              , \hspace{0.4cm} \mu^{\mathrm{t}} = {1- g_{a/\ell}\over 6\pi\eta a},
\end{eqnarray}

The passive translational and rotational mobility tensors of the spheres follow
from Eqs. (\ref{vel_with_ext}) and (\ref{omega_with_ext}), and they read as
$\Mb_{ii}^{\mathrm{\mathrm{tt}}} = M^{\mathrm{t}} \Ib$, $\Mb_{ii}^{\mathrm{rr}}
= M^{\mathrm{r}} \Ib$. Furthermore, for spheres, we have
$\Mb_{ii}^{\mathrm{tr}} = \Mb_{ii}^{\mathrm{rt}} = \mathbf{0}$ \cite{dhont}
Likewise, we can read off the active mobility tensors  from Eq.
(\ref{v_swim_final}) as $\mutens_{ii}^{\mathrm{tt}} = \mu^{\mathrm{t}} \Ib$ 
\tca{and from Eq. \eqref{omega_with_ext} as $\mutens_{ii}^{\mathrm{rt}} =
\mathbf{0}$.}
The remaining active self-mobilities, 
\tca{$\mutens_{ii}^{\mathrm{rr}}$ and $\mutens_{ii}^{\mathrm{tr}}$}, 
are irrelevant to the dynamics as the presented model does not include any
active torques.

\subsection{Cross mobility tensors} 

We now discuss the cross-mobility tensors $\Mb_{ij}^{\alpha \beta}$ and
$\mutens_{ij}^{\alpha \beta}$ for $i\ne j$ that result from the interactions
between two distinct swimmers $i$ and $j$. In particular, we derive the form of $\mutens_{ij}^{\alpha \beta}$  tensors.
To obtain the explicit form of the hydrodynamic interactions between the
swimmers, we exploit the Faxen's theorems  \cite{dhont}.

The Faxen theorems relate
the translational and angular velocities  of a sphere to the  forces,  torques
and external flows imposed on it. Hence, the translational and rotational
velocities of the $i$th swimmer  in  the flow generated by the motion of all
the other swimmers are given by
\begin{eqnarray}
\vb_i &=& M^{\mathrm{t}} \fb_i^{\mathrm{ext}} 
        + \mu^{\mathrm{t}} \fb_i^{\mathrm{sp}}  
        + \left(1 + {a^2\over 6} \nabla_{r_i}^2\right) \sum_{j\ne i} \ub_j(\rb_i), \label{v_i_full}\\
\omegab_i &=& M^{\mathrm{r}} \Lb_i^{\mathrm{ext}} + {1\over 2}\nablab_{r_i} \times \sum_{j\ne i} \ub_j (\rb_i),
\label{omega_i_full} 
\end{eqnarray}
where $\sum_{j\ne i} \ub_j(\rb_i)$ is the total flow field in the absence of
the $i$th swimmer at its center $\rb_i$ created by all the other swimmers.
Substituting the explicit form of   $\ub_j(\rb_i)$  from Eq.
(\ref{flow_individual}) in the above equations provides us with the mobility
tensors $\Mb_{ij}^{\alpha \beta}$ for $i\ne j$ \cite{dhont}: 
\begin{eqnarray}
\Mb^{\mathrm{tt}}_{ij}  &=& 
                M^{\mathrm{t}} \left[ \vphantom{\left({a\over r_{ij}}\right)^3} 
                {3 a\over 4 r_{ij}}(\Ib + \rhat_{ij}\rhat_{ij})\right. \nonumber \\
              && ~~~~~~  \left. + {1\over 2}\left({a\over r_{ij}}\right)^3
                  (\Ib -3 \rhat_{ij}\rhat_{ij}) \right] \equiv \Mb^{\mathrm{tt}} (\rb_{ij}), \\
\Mb^{\mathrm{rr}}_{ij}  &=& -M^{\mathrm{r}} {1\over 2} \left({a\over r_{ij}}\right)^3 
                                                    (\Ib -3 \rhat_{ij}\rhat_{ij}) 
                                                        \equiv \Mb^{\mathrm{rr}} (\rb_{ij}), \\
\Mb^{\mathrm{tr}}_{ij} &=& \Mb^{\mathrm{rt}}_{ij} 
                       = M^{\mathrm{r}} a \left({a\over r_{ij}}\right)^2 \rhat_{ij} \times
                                                        \equiv \Mb^{\mathrm{tr}} (\rb_{ij})\times.
\label{Mtensors}
\end{eqnarray}
in which $\rb_{ij} := \rb_i - \rb_j $ and  $r_{ij}:=|\rb_{ij}|$ and we have only
kept the  terms up to the order $(a/r_{ij})^3$.  The tensor
$\Mb^{\mathrm{tt}}_{ij}$ given above is the well known Rotne-Prager mobility
tensor for passive spheres. 

In a similar manner, we derive the active mobility tensors
$\mutens^{\mathrm{\alpha\beta}}_{i\ne j}$ correct up to the order $(a/r_{ij})^3$ as
given below: 
\begin{eqnarray}
\mutens^{\mathrm{tt}}_{ij} &=&  \mutens^{\mathrm{tt}}_0 (\rb_{ij}) 
                            + \mutens^{\mathrm{tt}}_\ell (\rb^\ell_{ij}) 
                            + \mutens^{\mathrm{tt}}_{\ell^*} (\rb^{\ell^*}_{ij}), \\
\mutens^{\mathrm{rt}}_{ij} &=&  \mutens^{\mathrm{rt}}_0 (\rb_{ij}) 
                            + \mutens^{\mathrm{rt}}_\ell (\rb^\ell_{ij}) 
                            + \mutens^{\mathrm{rt}}_{\ell^*} (\rb^{\ell^*}_{ij}),
\end{eqnarray}
where $\rb_{ij}^{\ell} := \rb_{ij} \pm \ell \nhat_j$, $\rb_{ij}^{\ell^*} :=
\rb_{ij} \pm \ell^* \nhat_j$.  The functions appearing on the R.H.S's are given by
\begin{eqnarray}
\mutens^{\mathrm{tt}}_0 (\rb) &=&  (1 - g_{a/\ell}) \Mb^{\mathrm{tt}} (\rb),  \\
\mutens^{\mathrm{tt}}_\ell (\rb) &=& - M^{\mathrm{t}} \left[{3 a\over 4 r}(\Ib + \rhat\rhat) 
                                                 + {1\over 4}\left({a\over r}\right)^3
                                                (\Ib -3 \rhat\rhat) \right], \\
\mutens^{\mathrm{tt}}_{\ell^*} (\rb) &=&  - g_{a/\ell}\, \mutens^{\mathrm{tt}}_\ell (\rb) \nonumber \\ 
                                            &&   + 2 j_{a/\ell} M^{\mathrm{t}} {3\over 4} \left({a\over r}\right)^3 
                                                    (\Ib -3 \rhat\rhat) \nonumber \\
                                            &&  + \tca{h_{a/\ell} \left[-\rhat +3 (\rhat\cdot\nhat_j)^2\rhat\right] 
                                                  M^{\mathrm{t}} {3\over 4} \left({a\over r}\right)^2 \nhat_j,} \\
\mutens^{\mathrm{rt}}_0 (\rb)          &=& (1 - g_{a/\ell}) \frac{1}{8\pi\eta} \frac{\rb}{r^3} \times, \\
\mutens^{\mathrm{rt}}_\ell (\rb)     &=& -\frac{1}{8\pi\eta} \frac{\rb}{r^3}\times, \\
\mutens^{\mathrm{rt}}_{\ell^*} (\rb) &=&  3 h_{a/\ell}\frac{\nhat_j\cdot\rb
                                    \left(\nhat_j\times\rb\right)\nhat_j}
                                     {8\pi\eta r^5}
                                  + g_{a/\ell} \frac{1}{8\pi\eta} 
                                                \frac{\rb}{r^3} \times. 
\end{eqnarray}
\tca{In the limit $a/\ell \rightarrow 0$,  we recover the active  mobility
tensors  of a bare point-force dipole:
\begin{eqnarray}
\label{eq:mobilitydipole}
 \mutens^{\mathrm{tt}}_{ij}&=&(1/8\pi\eta) [(\Ib + \rhat_{ij}\rhat_{ij})/r_{ij} -
(\Ib + \rhat^\ell_{ij}\rhat^\ell_{ij})/r^\ell_{ij}] \\ 
\mutens^{\mathrm{rt}}_{ij}&=& (1/8\pi\eta) [\rb_{ij}/r_{ij}^3 - \rb^\ell_{ij}/ 
(r^\ell_{ij})^3] \times .
\end{eqnarray}
}

The presented  swimmer model does not include any active torques. Thus, the
remaining active cross-mobilities $\mutens_{ij}^{\mathrm{tr}}$ and
$\mutens_{ij}^{\mathrm{rr}}$ are irrelevant  for the description of  the
dynamics.  Note that unlike the passive mobility tensors, the active
contributions depend  on the swimmers' orientations. In the next subsection, we
will discuss  the consequences of active forces on the hydrodynamic
interactions for finite-sized swimmers.

\subsection{ Activity-induced hydrodynamic interactions } 

\begin{figure*}
\begin{center}
\includegraphics[width=1.7\columnwidth]{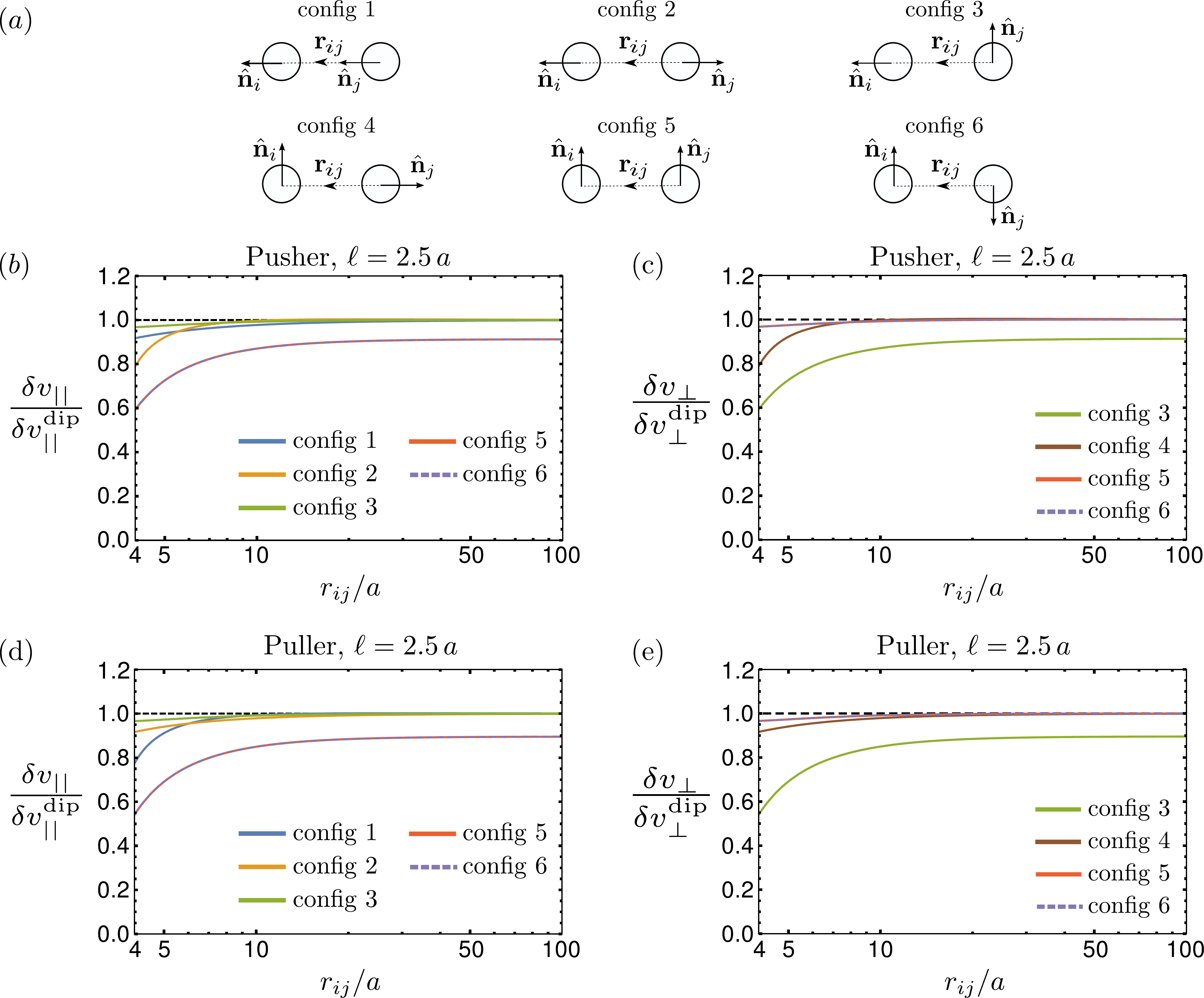}

\caption{The effect of the finite size of the swimmers on hydrodynamic
interactions.  (a) Schematics of six configurations for two swimmers $i$, $j$
with respective orientations $\nhat_i$, $\nhat_j$ and a separation vector
$\rb_{ij}$. \tca{(b)-(e) The ratios $\delta v_{||} / \delta v^{\rm{dip}} _{||}$ and
$\delta v_{\perp}/ \delta v^{\rm{dip}}_{\perp}$ plotted as a function of the separation $r_{ij}/a$ for $\ell = 2.5 a$. Here, $\delta v_{||}$ ($\delta v_{\perp}$) and $\delta v^{\rm{dip}} _{||}$ ($\delta v^{\rm{dip}} _{\perp}$)  
are the velocity increments  in  directions parallel (perpendicular)
to the orientation of a swimmer and its associated  point-force dipole  resulting from their  hydrodynamic 
interactions with another swimmer  and another force dipole, respectively.
 The plots for a pusher [(b) and (c)] and a puller [(d) and (e)] are shown
separately.}
\tca{Note that the plots for configurations $5$ and $6$ in all the sub-figures have
an exact overlap.}
}

\label{v_interaction}
\end{center}
\end{figure*}

We end our results with a preliminary examination of the consequences of active
forces on the hydrodynamic interactions and  we will pay a special attention to
the consequences of the swimmer's finite size.  For this purpose, we consider the
velocity increment $\tca{\delta \vb_i }= \mutens^{\mathrm{tt}}_{ij} \cdot
\fb^{\rm{sp}}_j$ induced on the swimmer $i$ due to the active forces $\pm
\fb^{\rm{sp}}_j$ of the swimmer $j$. Let, $\delta v_{||}$ and $\delta
v_{\perp}$  be the magnitudes of this velocity along and normal to $\nb_i$,
respectively. 
\tca{Likewise, we denote the corresponding speeds for an associated point-force
dipole of strength $S^{\mathrm{eff}}$ as $\delta v^{\rm{dip}}_{||}$ and $\delta
v^{\rm{dip}}_{\perp}$, respectively.  They result from the interaction between
the two corresponding associated  point-force dipoles. Their values are
obtained by using  the mobility tensor given in Eq. \ref{eq:mobilitydipole} evaluated
at $\rb_{ij} \to \rb_{ij} + (\ell - \ell^{\rm{eff}})\nhat_j$.}

In Fig.  \ref{v_interaction} we show the ratios $\delta v_{||}/ \delta
v^{\rm{dip}}_{||}$ and $\delta v_{\perp}/\delta v^{\rm{dip}}_{\perp}$ as a
function of the swimmer-distance $r_{ij}$ for $\ell = 2.5 a$ for a few typical
configurations of the swimmers $i$ and $j$ depicted schematically in Fig.
\ref{v_interaction} (a).  \tca{ We observe that for the configurations 5 and 6,
$\delta v_{||}/ \delta v^{\rm{dip}}_{||}$ depicted in Figs.
\ref{v_interaction} (b) and (d) and for the configuration 3,  $\delta
v_{\perp}/\delta v^{\rm{dip}}_{\perp}$  presented in Figs.  \ref{v_interaction}
(c) and (e) do not converge to unity even at  very large distances.  For all
other configurations, the ratios go  to unity for $r_{ij} > 20 a$
as one would expect.}

\tca{The disagreement between the activity-induced velocity increments of a
finite-sized swimmer and that of a dipolar swimmer for some of the configurations can
be understood as follows.  The far-distance expansion of the mobility tensor
for the swimmer yields to 
\begin{eqnarray}
\label{eq:farfieldj}
\delta \vb^{sp}(\rb_{ij}) &\equiv& \mutens^{\mathrm{tt}}_{ij} \cdot
                                \fb^{\rm{sp}}_j \nonumber \\
                      &=& \frac{1}{8\pi\eta} \frac{S^{\mathrm{eff}}}{r_{ij}^2} 
                          \left[-1 + 3 (\rhat_{ij}\cdot\nhat)^2\right]\rhat_{ij} 
                          + \mathcal{O}(\frac{1}{r_{ij}^3} ). \nonumber \\
                  \, 
\end{eqnarray}
As  expected, the leading order term above is identical to the corresponding
leading order term of the mobility tensor of the associated force dipole that
is centered at $\rb_{ij} + (\ell-\ell_{\mathrm{eff}}/2)\nhat_j$. However, the
coefficients of the next leading order term $\mathcal{O}(\frac{1}{r_{ij}^3})$
for the swimmer is different from that for the associated dipole.}

\tca{We note that the leading order term in Eq. \eqref{eq:farfieldj} is always
along $\rhat_{ij}$. For the specific configurations 5 and 6, $\hat{\nb}_i \perp
\rhat_{ij}$. Hence, the velocity increment parallel to $\hat{\nb}_i$ is
determined by the next leading order term  ($\propto 1/r_{ij}^3$) that differs
from that of force-dipoles. Thus,  the ratios $\delta v_{||}/ \delta
v^{\rm{dip}}_{||}$ for configurations 5 and 6 of Figs.  \ref{v_interaction} (b)
and (d)  do not converge to unity. Similarly, for the configuration 3, the
velocity increment perpendicular to $\nhat_i$, $\delta v_{\perp}$, is normal to
$\rhat_{ij}$ and its value is determined by the $\propto 1/r_{ij}^3$ term that
is different from that of the dipoles.  Therefore, $\delta v_{\perp}/\delta
v^{\rm{dip}}_{\perp}$ of the configuration 3 presented in Figs.
\ref{v_interaction} (c) and (e) also do not converge to unity.  These
observations highlight the importance of the swimmers finite-size  for their
hydrodynamic interactions. In our preliminary analysis, we have investigated
only few representative configurations. In a suspension,   the difference  in
the velocity increments $\delta \vb$ and $\delta \vb^{\rm{dip}}$ in general
depends on the relative orientations of the two swimmers with respect to their
interconnecting line.}

\section{Conclusions}
\label{sect:conclusions}

We have investigated the finite body-size  effects on the flow properties and
hydrodynamic interactions of microswimmers by considering a minimal model for
rigid-bodied microswimmers.  Notably, we have investigated the consequences of
the extended surface of the microswimmer on the interplay between its
self-propulsion and flow properties.  The model parameters include the body
size $a$ and the distance $\ell$ between the 
\tca{center of the body and the point of application of the thrust $-
\fb^{\mathrm{sp}}$ on the fluid.}
The presented model has important differences with the squirmer model for
microswimmers \cite{squirmer_lighthill1952, squirmer_blake1971, bickel2013}.
The latter applies to swimmers with deformable surfaces. Moreover, in contrast
to the squirmer model, the finite body-size and self-propulsion in this model
presents two separate length scales $a$ and $\ell$ intrinsic to each
individual. As a result, the flow field of this model at a separation $r$
crucially depends on both $\ell/r$ and $a/\ell$.

We have calculated the swimmer's self-propulsion velocity and  flow field as a
function of $a/\ell$.  Our analysis shows that the behavior of the minimal
swimmer deviates significantly from that of a point-force dipole unless
$a/\ell \to 0$.  The far field flow of  the swimmer can be mapped to that of a
point-force dipole with an effective dipolar strength given by
$S^{\mathrm{eff}}= f^{\mathrm{sp}}\ell^{\mathrm{eff}}$ where $\ell^{\rm{eff}}$
is smaller than 
\tca{Thus, we  have demonstrated that a point-force dipole  can provide a
relatively good description of the far filed behavior of a finite-sized spherical
swimmer provided that we use $\ell^{\rm{eff}}=\ell^{\rm{eff}} (\ell,a)$ as the effective
distance between the two point-forces.}

\tca{At intermediate distances, the flow field presents a remarkably different
angular dependency in comparison to that of a force dipole due to the
contribution of higher order multipoles.   Furthermore, the finite body-size
breaks the inherent front-back symmetry of the flow field of dipolar swimmers.
Therefore, the model naturally incorporates the head-tail asymmetry at the
hydrodynamic interaction level that is an intrinsic feature of many
microswimmers. These results highlight the necessity for  the inclusion of  the
finite body-size in simulations of microswimmer suspensions at moderate
concentrations.  }

To  investigate  the finite body-size  effects on hydrodynamic interactions and
on the collective dynamics of the swimmers, we have  derived the grand mobility
tensor for the presented  model.     A  preliminary investigation demonstrates
that the finite body-size  
\tca{significantly}
affects  the strength of hydrodynamic interactions. The consequences of the
finite body-size and particularly  the head-tail  asymmetry of the flow field
on the collective dynamics of swimmers is not yet probed.  The collective
behavior of minimal spherical swimmers can be investigated by Stokesian
dynamics simulations and will be a subject of a future study. The mobility
matrix elements describe long-range interactions and have rather cumbersome
forms.  Nevertheless, they can be properly decomposed by an  Ewald summation
technique  to account for the long-range of hydrodynamic interactions in
numerical simulations \cite{tapan_tobepublished1}.

To summarize,   our model captures the essential features of microswimmers with
finite-body size in a minimal fashion.  It opens the door for exploring
collective behavior of microswimmers at intermediate concentrations where
point-force dipole approximation fails.    The  presented model is general and
it can be used to model microwswimmers  that  have a rigid body with  flexible
propelling appendage by a suitable choice of parameters. 
 
\begin{acknowledgments}

\noindent\small{We are  grateful to J. F.  Brady for insightful discussions and
we acknowledge financial support from the German Science Foundation
(http://www.dfg.de) within SFB TRR 146 (http://trr146.de). }

\end{acknowledgments}


\begin{thebibliography}{56}
\bibitem [{zer()}]{zeroangularfoot}%
  \BibitemOpen
  \href@noop {} {}\bibinfo {note} {Angluar velocity of the swimmer generated by
  the flow $\ub_0(\rb)$ is $(1/2)\nablab_{r_{\mathrm{cm}}}\times
  \ub_0(\rb_{\mathrm{cm}})$ \cite{dhont}. This equates to zero, as can be seen
  by evaluating the curl, because the point force $\fb^{\mathrm{spc}} =
  -f^{\mathrm{sp}}\nhat$ points along a radial direction for the
  swimmer.}\BibitemShut {Stop}%
\bibitem [{int()}]{integration_oseen_foot}%
  \BibitemOpen
  \href@noop {} {}\bibinfo {note} {Here we use the result of the integration of
  the Oseen tensor over a sphere $\oint_S dS\Tb(\rb_S - \rb_S^\prime) = \Ib
  2a/3\eta$, where both $\rb_S$ and $\rb_S^\prime$ are points on the surface of
  the sphere of radius $a$ \cite{dhont}.}\BibitemShut {Stop}%
\end{thebibliography}

%

\end{document}